\documentclass[showpacs,amsmath,amssymb,showpacs,twocolumn,superscriptaddress, aps]{revtex4}
\usepackage{graphicx}
\usepackage{epstopdf}
\usepackage{bm}
\usepackage{amsmath}
\usepackage{amssymb}
\usepackage[raggedright,nooneline,large]{subfigure}
\usepackage{braket}
\usepackage[english]{babel}
\usepackage{color}
\usepackage{ulem}

\begin{document}
\title{Variational dynamics of the sub-Ohmic spin-boson model on the basis of multiple Davydov $\mathbf{D_{1}}$ states}


\author{Lu Wang}
\affiliation{Division of Materials Science, Nanyang Technological University, Singapore 639798, Singapore}
\affiliation{Department of Physics, Zhejiang University, Hangzhou 310027, China}
\author{Lipeng Chen}
\affiliation{Division of Materials Science, Nanyang Technological University, Singapore 639798, Singapore}
\author{Nengji Zhou}
\affiliation{Department of Physics, Hangzhou Normal University, Hangzhou 310046, China}
\author{Yang Zhao} \email{YZhao@ntu.edu.sg}
\affiliation{Division of Materials Science, Nanyang Technological University, Singapore 639798, Singapore}
\date{\today}

\begin{abstract}
Dynamics of the sub-Ohmic spin-boson model is investigated by employing a multitude of the Davydov $\mathrm{D}_{1}$ trial states,
also known as the multi-$\mathrm{D}_{1}$ \textit{Ansatz}. Accuracy in dynamics simulations is improved
significantly over the single $\mathrm{D}_{1}$ \textit{Ansatz}, especially in the weak system-bath coupling regime. The reliability of the
multi-$\mathrm{D}_{1}$  \textit{Ansatz} for various coupling strengths and initial conditions are also systematically
examined, with results compared closely with those of the hierarchy equations of motion and the path integral Monte Carlo approaches. In addition, a
coherent-incoherent phase crossover in the nonequilibrium dynamics is studied through the multi-$\mathrm{D}_{1}$ \textit{Ansatz}.
The phase diagram is obtained with a critical point $s_{c}=0.4$.  For $s_{c}<s<1$, the coherent-to-incoherent crossover occurs at a certain
coupling strength, while the coherent state recurs at a much larger coupling strength. For $s<s_{c}$, only the coherent phase exists.

\end{abstract}
\maketitle

\section{Introduction}

\begin{figure}[tbp]
  \centering
  \subfigure[\vspace{1cm}]{
    \label{fig:model}
    \includegraphics[clip]{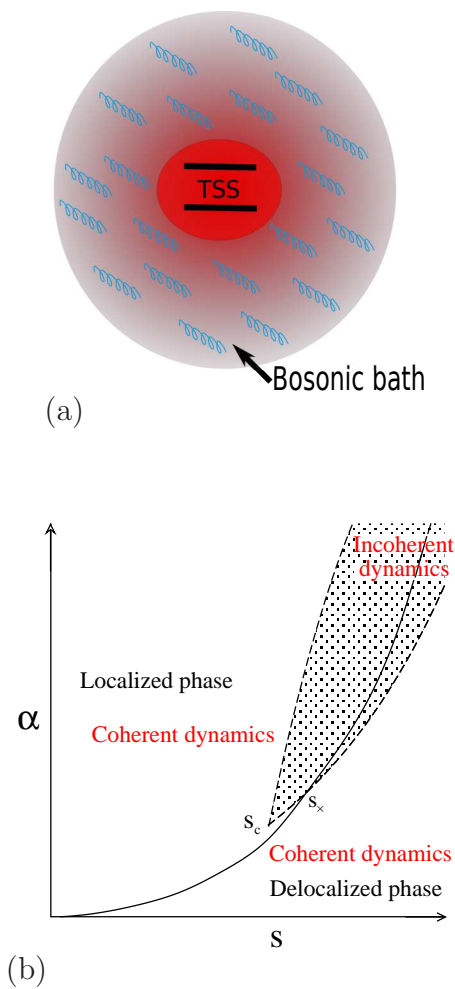}
  }
  \subfigure[]{
    \label{fig:phase_dia}
    \includegraphics[clip]{fig1b.eps}
  }
  \caption{(a)~Schematic of the SBM. (b)~Sketch of the SBM phase diagram. The solid line which is the critical
    line of the localized-delocalized phase transition, partitions the domain
    into the delocalized (below) and the localized (above) phase.  The two dashed lines are coherent-incoherent crossover lines.  The lower
    crossover line intersects with the phase transition line at $s_{\times}$ and the two crossover lines meet at the critical point $s_{c}$.
    $s_{c}$ separates the interval into two parts.  For
    $s_{c}<s<1$, the two dashed line separate the domain into three parts. The shaded area corresponds to the incoherent state and the other
    two are in the coherent state.  For $s<s_{c}$, there is only coherent state. }
  \label{fig:mod}
\end{figure}

It is of fundamental importance to study relaxation and coherence of open quantum systems attached to dissipative baths
\cite{Leggett,Weiss,Yao,Duan}, a paradigm of which is the spin-boson model (SBM), describing a two-level system coupled with a bath of
harmonic oscillators, as shown schematically in Fig.~\ref{fig:model}.  The SBM has garnered attention due to its wide-ranging applications in a
variety of physical situations, such as quantum computation \cite{Makhlin,Vion,Koch}, quantum phase transitions
\cite{Vojta,Alvermann,Winter,Lu,Zhao}, spin dynamics \cite{Leggett,Yao,Duan} and electron transfer in biological molecules
\cite{Marcus,Muhlbacher}.  The Hamiltonian of the SBM can be written as
\begin{equation}\label{hamilton}
  \hat{H}=\frac{\epsilon}{2} \sigma_{z}-\frac{\Delta}{2}\sigma_{x}
  +\sum_{l}\omega_{l}b_{l}^{\dagger}b_{l}
  +\frac{\sigma_{z}}{2}\sum_{l}\lambda_{l}\left(b_{l}^{\dagger}+b_{l}\right),
\end{equation}
where $\epsilon$ and $\Delta$ are the spin bias and the tunneling constant, respectively, $\sigma_{x}$ and $\sigma_{z}$ are Pauli
matrices, $b_{l}^{\dagger}$ ($b_{l}$) is the boson creation (annihilation) operator of the $l$th mode with frequency $\omega_l$, and
$\lambda_{l}$ labels the spin-boson coupling strength associated with the $l$th mode.  The environment and its coupling to the system are completely
characterized by a spectral density function $J(\omega)$:
\begin{equation}
  J(\omega)=\sum_{l}\lambda_{l}^{2}\delta{(\omega-\omega_{l})}=2\alpha\omega_{c}^{1-s}\omega^{s}e^{-\omega/\omega_{c}}
  \label{spectral_soft}
\end{equation}
where $\alpha$ is the dimensionless coupling strength, and $\omega_{c}$ denotes the cutoff frequency.  The bosonic Ohmic bath is specified by
$s=1$, and $s<1$ ($s>1$) denotes the sub-Ohmic (super-Ohmic) bath.  In the presence of an Ohmic bath, the SBM can be mapped onto the anisotropic Kondo model
using bosonization techniques \cite{Leggett}, and there exists a Kosterlitz-Thousless-type phase transition that
separates a non-degenerate delocalized phase from a doubly degenerate localized one, as well as a turnover from a coherent phase to
an incoherent one.  However, both the static and dynamic properties of the sub-Ohmic SBM are still under debate due to inherent difficulties in
treating low-frequency bath modes.

Compared to the Ohmic case, the sub-Ohmic SBM is characterized by much pronounced coupling to the low-frequency bath modes, which
makes it harder to describe the dynamics accurately as the low frequency modes generally lead to strongly non-Markovian dynamics and
long-lasting bath memory effects.  The sub-Ohmic SBM has been investigated by several sophisticated computational methods.  The numerical
renormalization group (NRG) approach developed by Wilson \cite{Bulla} seems to reveal a second order quantum phase transition
for $0<s<1$ together with weakly damped coherent oscillations on short time scales even in the localized phase in the deep sub-Ohmic regime,
i.e., $s\ll{1}$.  A continuous imaginary time cluster algorithm based on the quantum Monte Carlo (QMC) method has been recently proposed to
study quantum phase transitions in the sub-Ohmic regime \cite{Winter}. Contrary to the NRG results, the QMC critical exponents are found to
be classical, mean-field like in the deep sub-Ohmic regime.  An extension of the Silbey-Harris \textit{Ansatz} \cite{Harris} has been utilized
to reveal a continuous transition with mean-field exponents for $0<s<0.5$ \cite{Chin}.

On the SBM dynamics, a consensus has yet to emerge on a basic physical picture.
The phase diagram is shown schematically in Fig.~\ref{fig:phase_dia}.
It is believed that for $s \leqslant 1$, with increasing coupling strength $\alpha$, a dynamical
coherent-incoherent crossover takes place first, followed by a delocalized-to-localized transition at a larger $\alpha$, as revealed
by the multilayer multiconfiguration time dependent Hartree (ML-MCTDH) approach and the quantum master equation techniques
\cite{Lv07,Wang_Zheng09,Wang08,Wang10}.  As shown in Fig.~1(b), two transition lines meet at a critical exponent $s_{c}$, partitioning the
plane into
three parts for $s>s_{c}$.  The shaded area corresponds to the incoherent phase, and the rest of domain belongs to the coherent phase.
The solid line depicts the localized-delocalized phase transition and intersects with the coherent-incoherent transition line
at an exponent point $s_{\times}$ larger than $s_{c}$.   For an extended sub-Ohmic SBM including off-diagonal coupling, a similar phase diagram was found using an analysis of the response function \cite{Nalbach13}.
Using the non-interacting blip approximation (NIBA), $s_{c}=s_{\times}=0.5$ is obtained \cite{Kast13}.  It is conjectured that
for $s < s_{c}$, the coherent-incoherent crossover has larger $\alpha$ than the
localized-delocalized phase transition \cite{Nalbach10}.  Recently, real-time path integral Monte Carlo (PIMC) techniques show that below the critical point $s_{c}$, the
nonequilibrium coherent dynamics can persist even under strong dissipation \cite{Kast13}.  Reliable algorithms are needed to probe the
dynamics accurately in the both weak and strong coupling strength regimes.

The SBM is analogous to the one-exciton, two-site version of the Holstein molecular crystal model \cite{Holstein}, and it is well known that
the Davydov \textit{Ansatz} and its variants \cite{Zhao1,Zhao2,Zhao3,Zhao4} are successful trial states in treating both static and dynamic
properties of the Holstein polaron. In addition, the single Davydov $\mathrm{D}_{1}$ \textit{Ansatz} has been applied to probe the dynamics of
the sub-Ohmic SBM \cite{Wu}, and it is shown that, counterintuitively, even in the very strong coupling regime, quantum coherence features
manage to survive under the polarized bath initial condition, in agreement with the PIMC results \cite{Kast13}.  Very recently, a multitude of
Davydov $\mathrm{D}_{1}$ \textit{Ansatz} (the multi-$\mathrm{D}_{1}$ \textit{Ansatz}) has been successfully constructed to study the ground-state
properties of the sub-Ohmic SBM with simultaneous diagonal and off-diagonal coupling.  Much more accurate results near the quantum phase
transition point, in agreement with those from the methods of density matrix renormalization group and exact diagonalization, have been
obtained by the multi-$\mathrm{D}_{1}$ \textit{Ansatz} \cite{Zhou1,Zhou2} than those by the single $\mathrm{D}_{1}$ \textit{Ansatz}. However, how
the multi-$\mathrm{D}_{1}$ \textit{Ansatz} fares with the dynamics of the SBM with an arbitrary initial state remains an issue to be
addressed.

In this paper, a time-dependent version of the multi-$\mathrm{D}_{1}$ \textit{Ansatz} \cite{Zhou1} is adopted to formulate an accurate
description of dynamic properties of the sub-Ohmic SBM with various system-bath coupling strengths, based on the Dirac-Frenkel
time-dependent variational principle.  The reliability and robustness of the \textit{Ansatz} are systematically probed, in comparison with the
hierarchy equations of motion (HEOM) and the PIMC approaches. The rest of the paper is organized as follows.  The multi-$\mathrm{D}_{1}$
\textit{Ansatz} and relevant physical quantities for the SBM are introduced in Sec.~\ref{sec:METHODOLOGY}.  In Sec.~\ref{sec:NRESULTS}, we
present numerical results, examine the discretization parameters for convergent results and show the superiority of the
multi-$\mathrm{D}_{1}$ \textit{Ansatz}.  In Sec.~\ref{sec:DISSCUSSION}, the reliability of the multi-$\mathrm{D}_{1}$ \textit{Ansatz} are checked
by comparing results from our method to those from the HEOM and PIMC approaches.  Furthermore, The dynamical coherent-incoherent
crossover is investigated through the multi-$\mathrm{D}_{1}$ \textit{Ansatz}.   Finally, Conclusions are drawn in Sec.~\ref{sec:CONCLUSION}.

\section{METHODOLOGY}
\label{sec:METHODOLOGY}

\subsection{Dirac-Frenkel Variation}

The multi-$\mathrm{D}_{1}$ \textit{Ansatz} is an improved trial wave function as compared to the single $\mathrm{D}_{1}$ \textit{Ansatz}
\cite{Wu}.  The time-dependent version of the multi-$\mathrm{D}_{1}$ \textit{Ansatz} can be written as
\begin{eqnarray}
  \Ket{\mathrm{D}(t)}&=&\ket{+}\sum_{n=1}^{M}A_{n}(t)\exp\left(\sum_{l}f_{nl}(t)b_{l}^{\dagger}-\mathrm{{H.c.}}\right)\ket{0}_{\rm ph}
  \nonumber\\
  &+&\ket{-}\sum_{n=1}^{M}B_{n}(t)\exp\left(\sum_{l} g_{nl}(t)b_{l}^{\dagger}-\mathrm{{H.c.}}\right)\ket{0}_{\rm ph}, \nonumber\\
  \label{Dsfun}
\end{eqnarray}
where $\rm H.c.$ denotes the Hermitian conjugate, $\left|+\right\rangle$ ($\left|-\right\rangle$) stands for the spin-up (spin-down)
state, and $\left|0\right\rangle_{\rm ph}$ is the vacuum state of the boson bath.  $A_{n}(t)$ and $B_{n}(t)$ are variational parameters representing
the amplitudes in states $\left|+\right\rangle$ and $\left|-\right\rangle$, respectively, and $f_{nl}(t)$ and $g_{nl}(t)$ are the
corresponding phonon displacements, where $n$ and $l$ denote the $n$th coherent state and the $l$th effective bath mode, respectively.
$M$ is the multiplicity, and the multi-$\mathrm{D}_{1}$ \textit{Ansatz} is reduced to the Davydov $\mathrm{D}_{1}$ \textit{Ansatz} when $M=1$.

Adopting the Lagrangian formalism of the Dirac-Frenkel time-dependent variational principle,
one can derive equations of motion (EOM) for the variational parameters,
\begin{eqnarray}
&&  \frac{\rm{d}}{\rm{d}t} \left( \frac{\partial L} {\partial \dot{u}_{n}^{*}} \right) - \frac{\partial L}{\partial u_{n}^{*}} = 0,
\label{dirac_frenkel}
\end{eqnarray}
where $u_{n}^{*}$ denotes the complex conjugate of the variational parameters $u_{n}$, which can be $A_{n}$, $B_{n}$, $f_{nl}$, or
$g_{nl}$.  The readers are referred to Appendix \ref{sec:VAR_EOM} for the detailed derivation of EOM for the variational parameters.  The
Lagrangian $L$ in Eq.~(\ref{dirac_frenkel}) associated with the trial state $\ket{D\left(t\right)}$ is defined as%
\begin{eqnarray}
  \label{Lag}
  L&=&\frac{i}{2}\Braket{\mathrm{D}(t)|\frac{\overrightarrow{\partial}}{\partial t}|\mathrm{D}(t)}
  -\frac{i}{2}\Braket{\mathrm{D}(t)|\frac{\overleftarrow{\partial}}{\partial t}|\mathrm{D}(t)} \nonumber \\
  && - \Braket{\mathrm{D}(t)|\hat{H}|\mathrm{D}(t)}.
\end{eqnarray}

\subsection{Spectral Density Discretization}
\label{subsec:discretization}

The bath is completely characterized by the spectral density function through the parameters $\lambda_{l}$ in
Eq.~(\ref{hamilton}), labeling the coupling strength of the spin to the $l$th bath mode with frequency $\omega_{l}$.
Using a discretization procedure, $\lambda_{l}$ can be obtained from the spectral density function as shown in Eq.~(\ref{spectral_soft}).
There are various discretization methods to achieve $\lambda_{l}$.  The simplest procedure is the linear discretization \cite{Wu}.  One divides the frequency domain $[0,\ \omega_{\max}]$ into $N_{b}$
equal intervals $\Delta\omega$, in which $N_{b}$ is the number of effective bath modes and $\omega_{\max}=4\omega_{c}$ is the upper bound of
the frequency.  The lower bound of the procedure is $\omega_{\min}=\Delta\omega=\omega_{\max} / N_{b}$ and the $l$th
frequency is $\omega_{l}=l\Delta \omega$.  The parameter $\lambda_{l}^{2}$ can be expressed as
\begin{equation}
  \lambda_{l}^{2} = J(\omega_{l}) \Delta \omega,
\end{equation}
by calculating the integral of the spectral density Eq.~(\ref{spectral_soft}) over $\omega$,
\begin{equation}
  \sum_{l=1}^{N_{b}} \lambda_{l}^{2}= \int_{0}^{\infty} d\omega J(\omega) \approx \sum_{l=1}^{N_{b}} J(\omega_{l}) \Delta \omega.
\end{equation}
In the linear discretization procedure, the number of modes $N_{b}$ can effect the dynamics in two aspects.  Firstly,
the Poincar\'e recurrence time $T_{p}=2\pi/\Delta\omega$ is determined by $\omega_{\min}$.  If the number of modes is not large enough,
artificial recurrence occurs in the time period of interest.  Secondly, in the sub-Ohmic regime, the dynamics of the observables may be
sensitive to the low frequency modes.  The number of modes required may vary for different initial bath conditions, such as
the factorized and polarized initial baths (see details in Sec. \ref{sec:NRESULTS}).  Under the factorized bath initial condition, the
dynamics is insensitive to the number of phonon modes.  However, the influence of $N_{b}$ on the dynamics with the polarized bath is
considerable.  For the linear discretization, a proper $\Delta \omega$ is needed to guarantee that the recurrence time is longer than time
interval we are interested in and sufficiently low frequencies are sampled.  Thus, the simulation is reliable only when numerous modes are
sampled.  Over ten thousand modes are required in order to obtain sufficiently low $\omega_{\min}$ in the linear discretization scheme.
Therefore, this discretization method requires an enormous number of EOM to be solved, leading to CPU time and memory constraints,
as well as numerically unstabilities.

The sub-Ohmic SBM, characterized by the effect of coupling to the low frequency bath modes, has long-lasting bath memory effects.  It is
possible to employ another discretization method to circumvent the disadvantage of linear discretization and reach a compromise between
accuracy and computational efficiency by focusing on the low frequency domain.  A method termed the logarithmic discretization is extensively
employed by the NRG techniques to probe ground-state properties \cite{Bulla}.  One divides the frequency domain $[0,\ \omega_{\max}]$ into
$N_{b}$ intervals, $[\Lambda^{-(l+1)},\ \Lambda^{-l}]\omega_{\max}$ ($l=0$, $1$, $2$, $\cdots$, $N_{b}-1$).  The parameters $\lambda_{l}$
and $\omega_{l}$ in Eq.~(\ref{hamilton}) can then be obtained as
\begin{equation}\label{Lambda}
  \lambda_{l}^{2} = \int_{\Lambda^{-l-1}\omega_{\mathrm{max}}}^{\Lambda^{-l}\omega_{\mathrm{max}}} \mathrm{d}x\, J(x),
\end{equation}
and
\begin{equation}\label{Omega}
  \omega_{l} = \lambda_{l}^{-2} \int_{\Lambda^{-l-1}\omega_{\mathrm{max}}}^{\Lambda^{-l}\omega_{\mathrm{max}}} \mathrm{d}x\, J(x)x.
\end{equation}
The logarithmic discretization can easily characterize the spectral density function in the low frequency domain.  The relatively large
discretization parameters such as $\Lambda=2$ or $1.5$ are usually employed in the NRG method, because only the ground
state is of interest.  Nevertheless for the dynamics, the excited states are important as well.  Smaller $\Lambda$ should be used, to
sample sufficiently low frequencies and take into account the effect of the high frequencies.  For the small $\Lambda$, the logarithmic
discretization is an approximation to the linear discretization.  The relative weight between different $\lambda_{l}$ is distorted by the
partition of the integral.  The distortion can only be improved when $\Lambda \to 1$.  It is difficult to estimate the Poincar\'e
recurrence time under the logarithmic discretization.  The recurrence occurs even when $\omega_{\min}$ is sufficiently
small for a large $\Lambda$.  In accordance to the previous discussion, we will adopt a logarithmic discretization due to its computational
efficiency.  Convergence of the results must be carefully analyzed in each case.

\subsection{Observables}

From Eq.~(\ref{Dsfun}), the norm of the trial wave function can be calculated as
\begin{eqnarray} \label{normalization}
  \mathcal{N}(t) & = & \Braket{\mathrm{D}(t)|\mathrm{D}(t)}  \nonumber \\
    & = &  \sum_{n,u=1}^{M} \left[A_{n}^{*}A_{u}R(f_{n}^{*}, f_{u}) + B_{n}^{*}B_{u}R(g_{n}^{*}, g_{u}) \right],
\end{eqnarray}
where $R(f_{n}^{*}, g_{u})=\langle f_{n}|g_{u}\rangle$ is the Debye-Waller factor defined as
\begin{equation}
  \label{debye_waller}
  R\left(f_{n}^{*},g_{u}\right) \equiv
  \exp\left[\sum_{l}\left(f_{nl}^{*}g_{ul}-\frac{1}{2}\left|f_{nl}\right|^{2}-\frac{1}{2}\left|g_{ul}\right|^{2}\right)\right],
\end{equation}
with the coherent state
\begin{equation}
  |g_m\rangle=\exp\left[\sum_{l} g_{ml}(t)b_{l}^{\dagger}-\mathrm{{H.c.}}\right] \ket{0}_{\rm ph}.
\end{equation}

It is well known that, the norm of the wave function $\mathcal{N}(t)$ should be unity at any time $t$, a fact that can be used to test the accuracy of the variational dynamics.  In the SBM, physical observables of interest are,
\begin{equation}
  \label{av_spin_xyz}
  P_{i}(t) \equiv \braket{ \sigma_{i}} = \Braket{\mathrm{D}(t)| \sigma_{i} | \mathrm{D}(t)},\mbox { $i=x$, $y$, $z$.}
\end{equation}
Here, $P_{x}(t)$ and $P_{y}(t)$ represent the real and imaginary part of the coherence between the spin-up and spin-down states,
respectively, and $P_{z}(t)$ describes the population difference.
By substituting the trial wave function of Eq.~(\ref{Dsfun}) into Eq.~(\ref{av_spin_xyz}), these quantities can be easily derived
as
\begin{eqnarray}
  \label{av_spin_x}
  P_{x}(t) &=& \sum_{n,u=1}^{M} [A_{n}^{*}B_{u} R(f_{n}^{*}, g_{u})  +  B_{n}^{*}A_{u} R(g_{n}^{*}, f_{u})], \nonumber\\
  \label{av_spin_y}
  P_{y}(t) &=& -i \sum_{n,u=1}^{M} [A_{n}^{*}B_{u} R(f_{n}^{*}, g_{u}) -B_{n}^{*}A_{u} R(g_{n}^{*}, f_{u})], \nonumber\\
  \label{av_spin_z}
  P_{z}(t) &=& \sum_{n,u=1}^{M} [A_{n}^{*}A_{u} R(f_{n}^{*}, f_{u}) - B_{n}^{*}B_{u} R(g_{n}^{*}, g_{u})].
\end{eqnarray}

To investigate the entanglement between the spin and the bath, we introduce the von Neumann entropy \cite{Costi,Amico},
\begin{equation}
S_{\rm{v}-\rm{N}}=-\omega_{+}\ln\omega_{+}-\omega_{-}\ln\omega_{-},
\end{equation}
where
\begin{equation}
\omega_{\pm}=(1\pm\sqrt{P_{x}^{2}+P_{y}^{2}+P_{z}^{2}})/2
\end{equation}
are the eigenvalues of the reduced density matrix obtained by tracing the density matrix over the bath degrees of freedom.

\section{NUMERICAL RESULTS}
\label{sec:NRESULTS}

The dynamical behavior of the sub-Ohmic SBM is sensitive to the bath initial conditions
\cite{Kast13,Nalbach10,Wu}.  Two initial conditions are often considered for
the phonon displacements: one is the factorized initial condition corresponding to the phonon vacuum
state with $f_{nl}(0)=g_{nl}(0)=0$, and the other is the polarized initial condition corresponding
to a displaced-oscillator state with $f_{nl}(0)=g_{nl}(0)=-\lambda_{l}/2\omega_{l}$ \cite{Wu}.
The spin is assumed to initially occupy the up state $|+\rangle$, i.e., $A_1(0)=1$, $B_1(0)=0$ and
$A_n(0)=B_n(0)=0$ $(n\neq{1})$.  In order to avoid singularities, uniformly distributed noise within $[-\varepsilon,\varepsilon]$ is added to the initial spin amplitudes ($\varepsilon=10^{-4}$) and phonon displacements
($\varepsilon=10^{-2}$).  Results have been averaged for a sufficiently large sampling size such
that the errors induced by the noise are negligible.

 In some previous work, such as Ref.~\cite{Nalbach10}, ``weak coupling'' usually denotes values of $\alpha$ that fall below the localized-delocalized transition point $\alpha_{c}$ . In this work, however, we focus on the dynamics of the sub-Ohmic SBM, and we use ``weak coupling'' to refer to the parameter space of small $\alpha$ in which the observables, such as $P_{z}$, behave as those in an underdamped oscillator.

\subsection{Dynamical behavior of the sub-Ohmic SBM}
\label{subsec:Dynamics}

The bosonic bath acts as a damping mechanism and the coupling strength $\alpha$ parameterizes the strength of the damping.  Probing the
properties of the dynamical observables with varying $\alpha$ offers insights into the quantum system in a dissipative environment.
In this subsection, we investigate the time evolution of the
population difference $P_{z}(t)$, the coherence between $\ket{+}$ and $\ket{-}$ $P_{x}(t)$, and the von Neumann entropy $S_{\rm{v}-\rm{N}}(t)$
by using the multi-$\mathrm{D}_{1}$ \textit{Ansatz}.  In the SBM, the two energy levels $\ket{+}$ and $\ket{-}$ can be used to describe the
donor and acceptor states in electron transfer, and then $P_{z}(t)$ is the population difference.  $P_{x}(t)$ is related to
the off-diagonal part of the reduced density matrix and describes the tunneling between the two energy levels.
$S_{\rm{v}-\rm{N}}(t)$ characterizes the information flow between the system and the bath.

\begin{figure}[tbp]
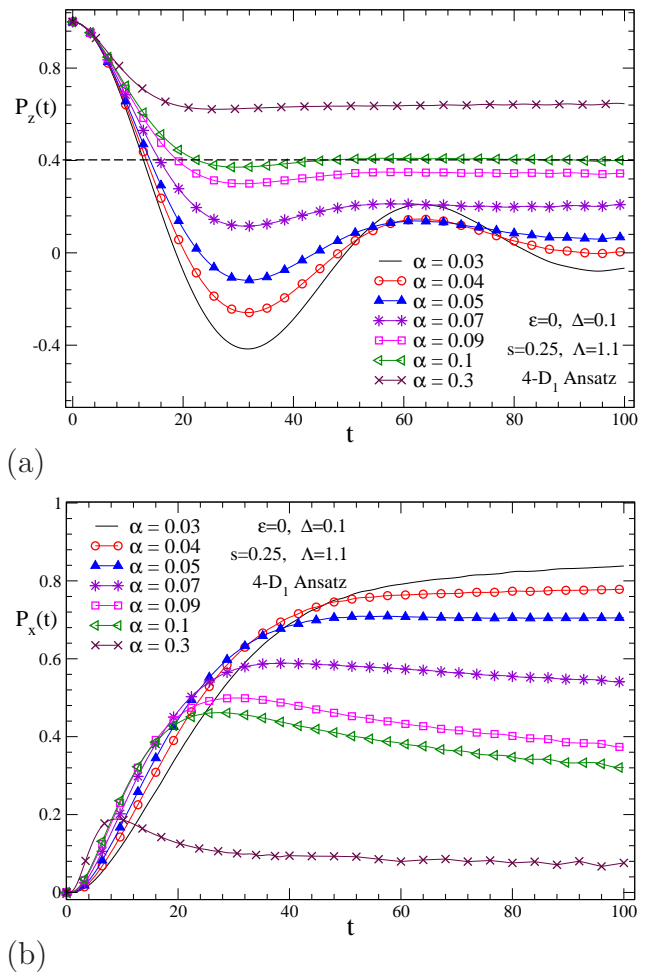

  \centering
  \subfigure[]{
    \label{fig:sigmaz_fa}
    \scalebox{0.35}{\includegraphics[clip]{fig2a.eps}}}
  \subfigure[]{
    \label{fig:sigmax_fa}
    \scalebox{0.35}{\includegraphics[clip]{fig2b.eps}}}
  \caption{(a)~The time evolution of the population difference $P_{z}(t)$, and (b)~the spin coherence $P_{x}(t)$ under a factorized bath
    initial condition are shown for various coupling $\alpha=0.03$, $0.04$, $0.05$, $0.07$, $0.1$ and $0.3$. Other Parameters used are $s=0.25$, $\Delta/\omega_{c}=0.1$, $\epsilon=0$, $M=4$ and $N_{b}=180$. The dashed line plotted on the steady-state value of $P_{z}(t)$ for $\alpha=0.1$, shows the
    existence of the weak oscillations for strong coupling.}
\end{figure}

In Fig.~\ref{fig:sigmaz_fa}, the population difference $P_{z}(t)$ is plotted for various coupling strengths $\alpha$ with the factorized
initial bath.  All coupling strengths in the simulations are above the localized-delocalized phase transition point
$\alpha_{c}=0.022$, thus the equilibrium value of $P_{z}(t)$, $P_{\mathrm{eq}}$, does not vanish.  With increasing
$\alpha$, $P_{\mathrm{eq}}$ approaches $1$.  In the weak coupling regime of $\alpha \in [0.03, 0.07]$, $P_{z}(t)$ oscillates and its
amplitude decays with time, as in a damped oscillator.  With increasing $\alpha$ in the range of $[0.03, 0.07]$, the bath
becomes more dissipative and the amplitude of $P_{z}(t)$ decreases, while the frequency of the oscillation almost does not change.  In the
strong coupling regime above $\alpha=0.07$, the amplitude of oscillation is suppressed severely by damping.  $P_{z}(t)$ quickly decays
to $P_{\mathrm{eq}}$ then holds a steady-state value for long times, similar to that in overdamped dynamics often represented as an
  exponential decay. As displayed in
Fig.~\ref{fig:sigmaz_fa}, for $\alpha=0.1$, however, a lobe highlighted by the dashed line exists below $P_{\mathrm{eq}}$, which deviates form the overdamped dynamics.   For larger $\alpha$, the lobe still
exists and is not marked for the small amplitude.  In addition, $P_{z}(t)$ reaches its minimum faster with increasing $\alpha$ implying the oscillation frequency
turns higher.  The higher frequency makes coherence possible before being suppressed by the
decay.  The phenomenon of coherent dynamics surviving under arbitrary large coupling for $s<1/2$, is first uncovered by the PIMC
method using the dynamics of $P_{z}(t)$ with $s=0.25$ and the polarized initial bath as an example \cite{Kast13}.  In the
Sec.~\ref{subsec:crossover}, we will discuss in detail such sustained coherence.

To probe the dynamics of the coherence between $\ket{+}$ and $\ket{-}$, $P_{x}(t)$ is calculated for $s=0.25$ with the
factorized initial bath.  Initially, $P_{x}(t)$ increases rapidly with time for all coupling strengths with a larger slope for stronger coupling.
A larger $\alpha$ leads to more rapidly evolving system dynamics,
thus the slope of $P_{x}(t)$ for the short times increases with increasing $\alpha$.  For long times,
the dynamical behaviors vary depending on the coupling strength.  For strong coupling,
$\alpha>0.07$, $P_{x}(t)$ first reaches its maximum before leveling off gradually.  The stronger the coupling is, the faster $P_{x}(t)$
reaches its maximum.  And the maximum of $P_{x}(t)$ is depressed by increasing $\alpha$.  For ultra-strong coupling
$\alpha=0.3$, after a decrease from its maximum, $P_{x}(t)$ holds a steady value.  Strong coupling shortens the oscillation time scale of
$P_{x}(t)$, a phenomenon also observed in the dynamics of $P_{z}(t)$.  The coherence between the two states is destructed under the impact of the bath for long times.
For weak coupling, $\alpha\in[0.03,0.07]$, after a transient stage, $P_{x}$ reaches a saturated value.  The dynamics for longer times is unknown and will be probed in further investigations.

\begin{figure}[tbp]
  \centering
  \subfigure[]{
    \label{fig:entropy_f}
    \scalebox{0.35}{\includegraphics[clip]{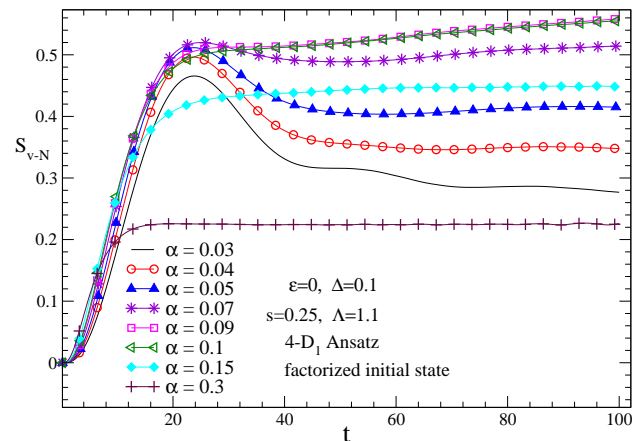}}}
  \subfigure[]{
    \label{fig:entropy_p}
    \scalebox{0.35}{\includegraphics[clip]{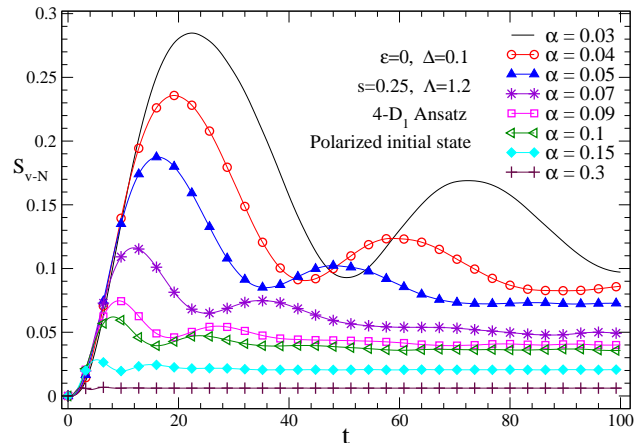}}}
  \caption{The time evolution of the von Neumann entropy $S_{\rm v-N}$ under (a)~the factorized bath initial condition, and (b)~the polarized
    bath initial condition for $\alpha=0.03$, $0.04$, $0.05$, $0.07$, $0.09$, $0.1$, $0.15$ and $0.3$. Other parameters are $s=0.25$,
    $\Delta/\omega_{c}=0.1$, $\epsilon=0$ and $M=4$. }
\end{figure}

To characterize the entanglement between the spin and bath, the von Neumann entropy $S_{\rm{v}-\rm{N}}(t)$ is plotted in
Fig.~\ref{fig:entropy_f} for the factorized initial bath.  It is found that $S_{\rm{v}-\rm{N}}(t)$ is quickly damped to a steady-state value, especially for the coupling strength $\alpha$ above $0.05$.  Though the steady values of $P_{z}(t)$ and $P_{x}(t)$ decrease
monotonically with the increasing coupling strength, those of $S_{\rm{v}-\rm{N}}(t)$ are nonmonotonic function of $\alpha$ and the
maximum steady-state value occurs at $\alpha \approx 0.09$, slightly different from $\alpha \approx 0.07$ as obtained by the
single $\mathrm{D}_{1}$ \textit{Ansatz} \cite{Wu}.
And our results show that, for long times, $S_{\rm{v}-\rm{N}}(t)$ reaches steady-state values, in disagreement with the conclusion in
Ref.~\cite{Kast13} that it will tend to zero in the strong coupling regime.  To clarify
this contradiction, simulations are performed with the same polarized initial bath as in Ref.~\cite{Kast13} for
$S_{\rm{v}-\rm{N}}(t)$, and results are shown in Fig.~\ref{fig:entropy_p}.  It is found that the entropy oscillates even when the coupling is
very strong and such a dynamical behavior is consistent with the coherent $P_{z}(t)$ found in the very strong coupling regime (see Fig.~\ref{fig:pimc}).  For the long times and with strong coupling, $S_{\rm{v}-\rm{N}}(t)$ also retains a
steady-state value which decreases monotonically with increasing $\alpha$.  Thus, it is our guess that the time interval studied in
  Ref.~\cite{Kast13} may be too short to observe stable values of $S_{\rm{v}-\rm{N}}(t)$.
  For the sub-Ohmic SBM, $S_{\mathrm{v-N}}(t)$ may approach zero very slowly, and the asymptotic behavior is beyond the reach of our approach as employed in this work.

\subsection{Discretization parameters}
\label{subsec:convergence}

If not carefully carried out, simulation of the SBM dynamics can be artificially influenced by
parameters that include the logarithmic discretization parameter $\Lambda$ and the frequency lower
bound $\omega_{\min}$.  In this subsection, a careful convergence test is performed for $\Lambda$ and $\omega_{\min}$.

\begin{figure}[tbp]
  \centering
  \scalebox{0.35}{\includegraphics[clip]{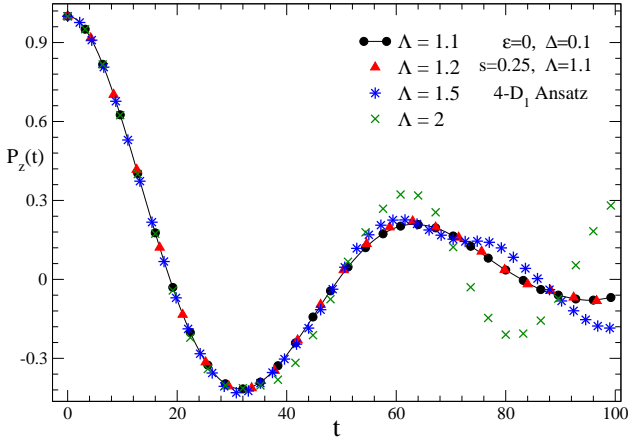}}
  \caption{The population difference $P_{z}(t)$ obtained by the multi-$\mathrm{D}_{1}$ \textit{Ansatz} with discretization parameters
    $\Lambda=1.1$, $1.2$, $1.5$ and $2$ are compared. The factorized bath initial condition is employed.  Other parameters are $s=0.25$,
    $\alpha=0.03$, $\Delta/\omega_{c}=0.1$ and $\epsilon=0$. }
  \label{fig:comB}
\end{figure}

To check the dependence of the computation convergence on $\Lambda$, $P_{z}(t)$ is simulated for $\alpha=0.03$,
$\omega_{\min}\approx10^{-7}$, $\Lambda=2$, $1.5$, $1.2$ and $1.1$, as shown in Fig.~\ref{fig:comB}.  For short times, dynamics computed
with different values of $\Lambda$ are nearly identical, but distinguishable deviations can be found at longer times.  For the curves with
$\Lambda=2$ and $1.5$, there exist high frequency oscillations which are an artifact as the system
cannot afford such high-energy oscillations.  While for the curves with $\Lambda=1.2$ and $1.1$,
the oscillations disappear and the difference between the two curves is negligibly small, which shows that the results are convergent
for these values of $\Lambda$.  As mentioned in Sec.~\ref{subsec:discretization}, the Poincar\'e recurrence time $T_{p}$ is difficult
to estimate using the logarithmic procedure.  Shown by our simulations,
with nearly identical $\omega_{\min}$, the recurrence occurs for larger $\Lambda$, i.e., $\Lambda=2$ and $1.5$.  We conclude that $T_{p}$ is also dependent on $\Lambda$ using the logarithmic discretization. In order to obtain reliable
results, a smaller $\Lambda$ should be chosen until the artificial oscillations disappear.

\begin{figure}[tbp]
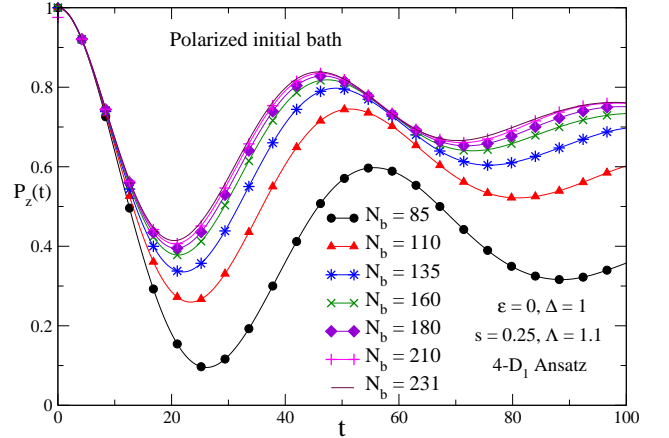

  \subfigure[]{
    \label{fig:com_mnum_fa}
    \scalebox{0.35}{\includegraphics[clip]{fig5a.eps}}}
  \subfigure[]{
    \label{fig:com_mnum_Po}
    \scalebox{0.35}{\includegraphics[clip]{fig5b.eps}}}
  \caption{The population difference $P_{z}(t)$ obtained by the multi-$\mathrm{D}_{1}$ \textit{Ansatz} with various mode numbers $N_{b}$ are
    compared. (a)~The factorized bath initial condition, and (b)~the polarized bath initial condition are employed. The discretization
    parameter $\Lambda=1.1$. Other parameters are $s=0.25$, $\alpha=0.03$, $\Delta/\omega_{c}=0.1$ and $\epsilon=0$.  }
  \label{fig:com_mnum}
\end{figure}

To probe the proper frequency lower bound $\omega_{\min}$, we perform simulations for $P_{z}(t)$ for various numbers of modes $N_{b}$
with $\alpha=0.03$ and both the factorized and polarized initial baths, and results are shown in Fig.~\ref{fig:com_mnum}.
We choose $\omega_{\min} \approx 10^{-3}$, $10^{-4}$, $10^{-5}$, $10^{-6}$, $10^{-7}$ and $10^{-8}$, which corresponds to $N_{b}=85$, $110$,
$160$, $180$, $210$ and $231$, respectively, to perform the simulations with $\Lambda=1.1$. In Fig.~\ref{fig:com_mnum_fa}, the results
with the
factorized initial bath are plotted.  All curves are found to coincide with each other, demonstrating that with the factorized initial bath,
the dynamics is insensitive to the low frequency modes.  However, with the polarized initial bath, the results vary
with the number of modes $N_{b}$ drastically, as shown in Fig.~\ref{fig:com_mnum_Po}. Convergence is reached only after the frequency lower bound is reduced to
$\omega_{\min}\approx 10^{-7}$. The dependence of $P_{z}(t)$ on $\omega_{\min}$ can be explained by a dynamical asymmetry,
i.e., a time-dependent bias, coming from the coupling term $\sigma_{z}/2\sum_{l}\lambda_{l}(b_{l}^{\dagger}+b_{l})$ in the Hamiltonian of
Eq.~(\ref{hamilton}).  As revealed by Ref.~\cite{Nalbach10}, the system first relaxes to a quasi-equilibrium state as determined by the
dynamical asymmetry, and then the asymmetry decays very slowly.  The average position of $P_{z}(t)$ is dominated by the low-frequency phonons.
Thus, the $\omega_{\min}$ must be small enough to make the contribution of the modes below $\omega_{\min}$ to the coupling term negligibly
small.  Our results show that the frequency lower bound  $\omega_{\min}$ yielding convergence with the factorized bath may be different
from that with the polarized bath even for the same coupling strength.  The convergence test for the number of modes must be performed
for each type of initial bath conditions.

\subsection{Accuracy of the multi-$\mathbf{D}_{1}$ \textbf{\textit{Ansatz}}}

\label{subsec:superiority}

The multi-$\mathrm{D}_{1}$ \textit{Ansatz} is an extension of the Davydov \textit{Ansatz}.  It is known
that the Davydov single $\mathrm{D}_{1}$
\textit{Ansatz} performs well in the strong and intermediate coupling regimes \cite{Yao}.  It may be less accurate in the weak coupling regime.
In this subsection, we will compare the results from the multi-$\mathrm{D}_{1}$ \textit{Ansatz} with those from the single $\mathrm{D}_{1}$ \textit{Ansatz} in the both weak and strong coupling regimes.

\begin{figure}[tbp]
  \centering
  \scalebox{0.35}{\includegraphics{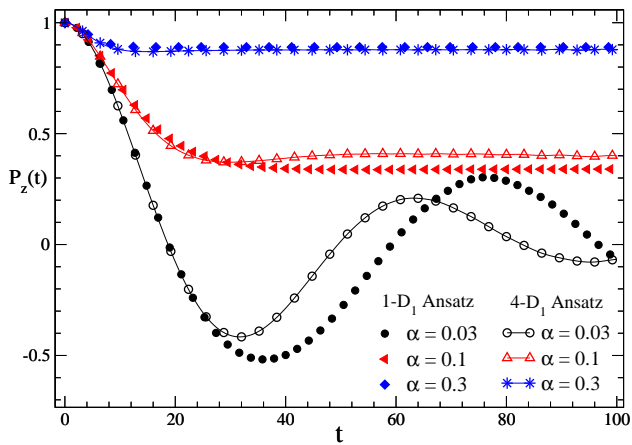}}
  \caption{Results of population difference $P_{z}(t)$ under the factorized bath initial condition from the single $\mathrm{D}_{1}$
    \textit{Ansatz} ($M=1$) and multi-$\mathrm{D}_{1}$ \textit{Ansatz} with $M=4$, are compared for the coupling strengths $\alpha=0.03$,
    $0.1$ and $0.3$.  Other parameters are $s=0.25$, $\Delta/\omega_{c}=0.1$, $\epsilon=0$, $\Lambda=1.1$ and $N_{b}=180$. }
  \label{fig:comD1}
\end{figure}

Figure~\ref{fig:comD1} shows the time evolution of the population difference $P_{z}(t)$ calculated by the single $\mathrm{D}_{1}$ and the
multi-$\mathrm{D}_{1}$ {\it Ans\"atze} under the factorized initial condition.  In the weak coupling regime (such as $\alpha=0.03$),
both the amplitudes and oscillation frequencies of $P_{z}(t)$ from the multi-$\mathrm{D}_{1}$ \textit{Ansatz} deviate significantly from those
calculated by the single $\mathrm{D}_{1}$ \textit{Ansatz}.  This is because the phonon wave function exhibits plane-wave like behavior in the
weak coupling regime, thus more phonon coherent states are needed to capture the accurate dynamics.  For strong coupling cases of $\alpha=0.1$ and
$0.3$, the single $\mathrm{D}_{1}$ \textit{Ansatz} still has room for improvements, though the difference
between the two curves is very small. At $s=0.25$, the single $\mathrm{D}_{1}$ \textit{Ansatz} only displays incoherent dynamics when
$\alpha \geqslant 0.1$.  While according to
Ref.~\cite{Kast13}, for the SBM in the deep sub-Ohmic regime, the coherence survives for any large coupling
strength.  These results indicate that the multi-$\mathrm{D}_{1}$ \textit{Ansatz}, as an improved trial wave function, can treat both
weak and strong coupling in a unified manner and captures more accurately dynamic properies.

\begin{figure}[tbp]
  \scalebox{0.35}{\includegraphics[clip]{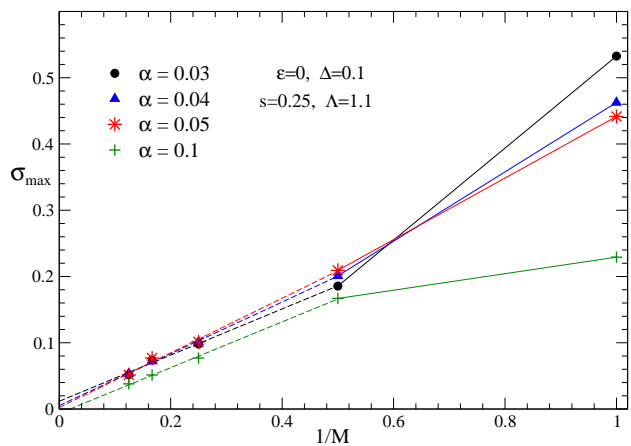}}
  \caption{The maximum value of the relative deviation $\sigma_{\max}(M)$ as a function of the multiplicity $M$ of the $\mathrm{D}_{1}$ {\it
      Ansatz} for the coupling strengths $\alpha=0.03$, $0.04$, $0.05$ and $0.1$.  An asymptotic value of
    $\sigma_{\max}(M\to\infty) \approx 0$ is obtained from the linear fitting (dashed lines).  The solid lines are used to connect data
    points with the same $\alpha$.  }
    \label{fig:errD1}
\end{figure}

To check the validity of the trial \textit{Ansatz}, we define the relative deviation $\sigma(M, t)$ for the trial state
$\ket{\mathrm{D}(M, t)}$,
\begin{equation}
  \label{rerror}
  \sigma(M, t)=\sqrt{ \Braket{\delta\left(M, t\right)|\delta\left(M, t\right)}}  / \bar{E}_{\mathrm{bath}},
\end{equation}
where $\Ket{\delta(M, t)}$ is the deviation vector quantifying how
faithfully $\ket{\mathrm{D}(M, t)}$ follows the Schr\"odinger equation,
\begin{equation}
  \Ket{\delta(M, t)} = \left( i\frac{\partial}{\partial t} - \hat{H} \right) \Ket{\mathrm{D}(M, t)},
\end{equation}
and $\bar{E}_{\mathrm{bath}}$ is the average energy of the bath within the simulation time.
Generally speaking, the smaller $\sigma(M, t)$ as a function of the multiplicity $M$, the more accurate the trial state
$\Ket{\mathrm{D}(M,t)}$ is.
Shown in Fig.~\ref{fig:errD1} for four
coupling strengths, $\alpha=0.03$, $0.04$, $0.05$ and $0.1$, are the maximum values of $\sigma(M,t)$, $\sigma_{\max}$, during the simulation
time period. For $M=1$, $\sigma_{\max}$ for $\alpha=0.1$ is much smaller than those for the weak coupling cases of $\alpha=0.03$, $0.04$ and $0.05$, which shows the single
$\mathrm{D}_{1}$ \textit{Ansatz} is more accurate for strong coupling.  Comparatively, $\sigma_{\max}$ for the multi-$\mathrm{D}_{1}$ \textit{Ansatz},
are much smaller than those for the single $\mathrm{D}_{1}$ \textit{Ansatz}.
To probe the asymptotic value of $\sigma_{\max}$ for $M \to \infty$, linear fitting
\begin{equation}
  \sigma_{\max}(M)=a_{1}M^{-1} + a_{0}\ (M>1),
\end{equation}
is employed, plotted as dashed lines in Fig.~\ref{fig:errD1}.  $\sigma_{\max} \to 0$ linearly with $1/M \to 0$.
The vanishing of $\sigma_{\max}(M)$ indicates that the multi-$\mathrm{D}_{1}$ \textit{Ansatz} improves the
variational dynamics of the SBM dramatically and may be numerical exact in the weak coupling regime in which the single $\mathrm{D}_{1}$
\textit{Ansatz} is known to be inaccurate.

\begin{figure}[tbp]
  \scalebox{0.35}{\includegraphics[clip]{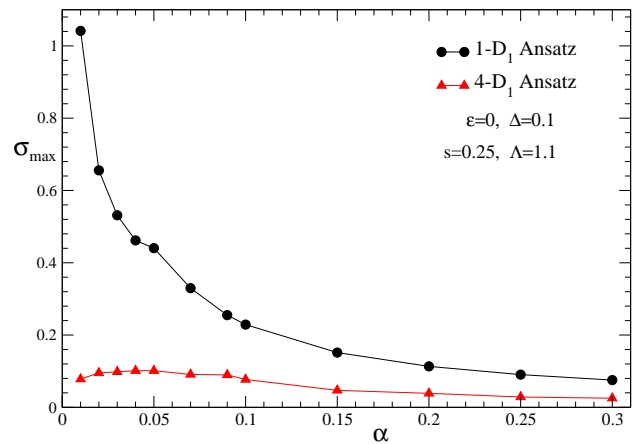}}
  \caption{The maximum value of the relative deviation $\sigma_{\max}(\alpha)$ as a function of the coupling strength $\alpha$ of the
    $\mathrm{D}_{1}$ \textit{Ansatz}.  The multiplicity $M=1$ and $4$ are plotted.  Other parameters $s=0.25$, $\Delta/\omega_{c}=0.1$,
    $\epsilon=0$, $\Lambda=1.1$ and $N_{b}=180$.}
    \label{fig:err_alpha}
\end{figure}

To probe whether the results are reliable for various coupling strengths, the maximum of the relative deviation $\sigma_{\max}$ is plotted for
$M=1$ and $M=4$ in Fig.~\ref{fig:err_alpha} as a function of $\alpha$.  In the weak coupling regime, from $\alpha=0.01$ to
$0.07$, calculated values of $\sigma_{\max}$ from the single $\mathrm{D}_{1}$ \textit{Ansatz} shown as solid dots are larger.  In the ultra-weak coupling regime, such
as $\alpha \approx 0.01$ and $0.02$, $\sigma_{\max}$ even exceeds $0.5$, which means that the results are unreliable.
The plane-wave like pattern of our results renders the single $\mathrm{D}_{1}$ \textit{Ansatz} invalid in this coupling regime.   As
displayed with solid triangles, results from the multi-$\mathrm{D}_{1}$ \textit{Ansatz} have much smaller $\sigma_{\max}$ than those from the single $\mathrm{D}_{1}$
\textit{Ansatz}, indicating that the multi-$\mathrm{D}_{1}$ \textit{Ansatz} can capture the plane-wave like pattern more effectively.  In the
strong coupling regime, the difference in $\sigma_{\max}$ between the single $\mathrm{D}_{1}$ and multi-$\mathrm{D}_{1}$ \textit{Ansatz}
narrows but $\sigma_{\max}$ for $M=4$ is still much smaller than that for $M=1$.  It is well known that the single $\mathrm{D}_{1}$
\textit{Ansatz} works well in the strong coupling regime \cite{Zhao1}.  As shown in Fig.~\ref{fig:err_alpha}, the multi-$\mathrm{D}_{1}$
\textit{Ansatz} preserves the advantage of the single $\mathrm{D}_{1}$ \textit{Ansatz} and performs even better in this regime.
It is found that $\sigma_{\max}(\alpha)$ for $M=4$ is very small for all coupling strengths, thus, it can be concluded that the
improved trial wave function can provide a more reliable description of the dynamical behavior of the SBM in the whole parameter regime.

\section{Discussion}
\label{sec:DISSCUSSION}

In this section we will discuss the dynamical coherent-incoherent crossover.  To confirm our method is reliable, results from the
multi-$\mathrm{D}_{1}$ \textit{Ansatz} are first compared with those from alternative numerical methods.  Then simulations are performed to
estimate the critical point $s_{c}$ and sketch the phase diagram.

\subsection{Benchmarking}
\label{subsec:Benchmarking}

\begin{figure}[tbp]
  \centering
  \scalebox{0.35}{\includegraphics[clip]{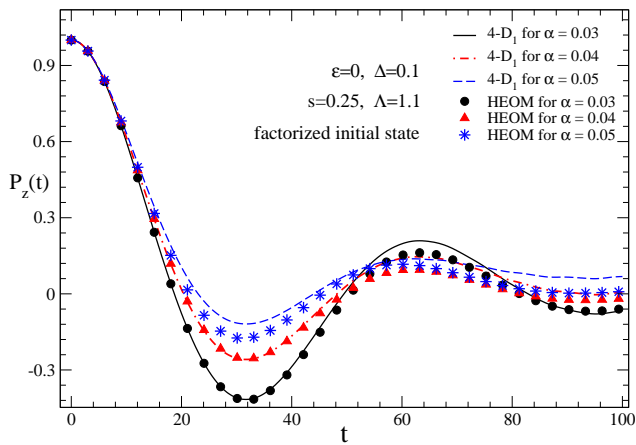}}
  \caption{Under the factorized bath initial condition, the population difference $P_{z}(t)$ calculated by the multi-$\mathrm{D}_{1}$ {\it
      Ansatz} with $M=4$ and the extended HEOM method using fitting technique for coupling strengths $\alpha=0.03$, $0.04$ and $0.05$ is
    displayed.  Other parameters are $s=0.25$, $\Delta/\omega_{c}=0.1$, $\epsilon=0$, $\Lambda=1.1$ and $N_{b}=180$. }
  \label{fig:heom2}
\end{figure}

To confirm that the multi-$\mathrm{D}_{1}$ \textit{Ansatz} is reliable under the factorized bath initial condition, we compare the population
difference $P_{z}(t)$ from the multi-$\mathrm{D}_{1}$ \textit{Ansatz} with that from the HEOM approach \cite{Ishizaki,Tanimura}, as
shown in Fig.~\ref{fig:heom2}.  By fitting the correlation function as a set of exponential functions, we construct the HEOM that can treat
dynamics of the SBM at zero temperature (see Appendix \ref{sec:HEOM} for details).  For the weak
coupling cases ($\alpha=0.03$ and $0.04$), $P_{z}(t)$ obtained from the multi-$\mathrm{D}_{1}$ \textit{Ansatz} agrees well with that from the HEOM
approach. The minor difference in amplitude at longer times may be due to the fact that the upper limit of the integral
for the correlation function, $C(t)$ in Eq.~(\ref{CF}), is infinity in the HEOM method, while in the multi-$\mathrm{D}_{1}$ \textit{Ansatz} the
upper limit of Eqs.~(\ref{Lambda}) and (\ref{Omega}) is set to $\omega_{\mathrm{max}} = 4\omega_{c}$.
For a stronger coupling strength, $\alpha=0.05$, results from these two methods exhibit behaviors of a damped oscillator with the same frequency.
However, distinguishable difference in the amplitudes of $P_{z}(t)$ can be seen.  The deviations may be attributed to that at stronger
coupling, the HEOM method needs more hierarchy equations to achieve convergence and the exponential fitting of the correlation
function in the HEOM method may not be accurate for larger $\alpha$.  In addition, the relative error $\sigma_{\max}$ at $\alpha=0.05$ is
as small as those at $\alpha=0.03$ and $0.04$, indicating that the variational dynamics based on the multi-$\mathrm{D}_{1}$  \textit{Ansatz} is
more accurate than that from the HEOM method at $\alpha=0.05$.

\begin{figure}[tbp]
  \centering
  \scalebox{0.35}{\includegraphics[clip]{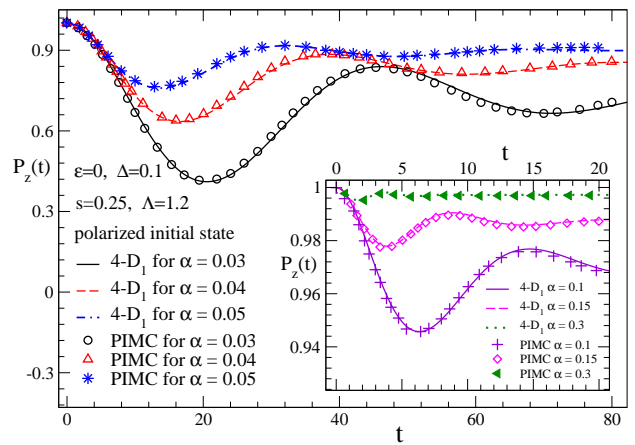}}
  \caption{The population difference $P_{z}(t)$ obtained by the multi-$\mathrm{D}_{1}$ \textit{Ansatz} are compared with those calculated by
    the PIMC method (extracted from Ref.\cite{Kast13}) under the polarized bath initial condition.  Other parameters used are $s=0.25$,
    $\Delta/\omega_{c}=0.1$, $\epsilon=0$, $\Lambda=1.2$ and $N_{b}=120$. }
  \label{fig:pimc}
\end{figure}

The polarized initial condition is especially of interest as it corresponds to the
typical experimental scenarios \cite{Weiss} in which the system is initialized in the ground state of
$\left. \sum_{l}\omega_{l}b_{l}^{\dagger}b_{l} +\frac{\sigma_{z}}{2}\sum_{l}\lambda_{l}\left(b_{l}^{\dagger}+b_{l}\right) \right|_{\sigma_{z}=1}$.
To verify the multi-$\mathrm{D}_{1}$ \textit{Ansatz} is reliable with the polarized initial
bath, the population difference $P_{z}(t)$ from the multi-$\mathrm{D}_{1}$ \textit{Ansatz} is compared with that from the PIMC method
\cite{Kast13}.  The comparison also checks the reliability of the multi-$\mathrm{D}_{1}$ \textit{Ansatz} in a broader parameter space of
$\alpha$.  As shown in Fig.~\ref{fig:pimc}, for both weak coupling cases of $\alpha=0.03$, $0.04$, $0.05$ and strong coupling cases of
$\alpha=0.1$, $0.15$, $0.3$, the results obtained by the multi-$\mathrm{D}_{1}$ \textit{Ansatz} are in excellent agreement with those by the
PIMC method.  The oscillation frequency turns higher and $P_{\mathrm{av}}$, the average position of $P_{z}(t)$, approaches $1$ with
increasing coupling.  As pointed out in Sec.~\ref{subsec:convergence}, the initial displacement $-\lambda_{l}/2\omega_{l}$ \cite{Wu}
introduces a time-dependent bias $\epsilon(t)$, and $\epsilon(t)$ decays very slowly.  Increasing $\alpha$ enlarges $\epsilon(t)$ and
moves $P_{\mathrm{av}}$ toward $1$.  Meanwhile, $\epsilon(t)$ also raises the oscillation frequency, and renders more oscillations to appear under
the influence of the polarized initial bath.

\subsection{Dynamical crossover}
\label{subsec:crossover}

In this subsection, we investigate the dynamical crossovers in the sub-Ohmic SBM.  According to Fig.~\ref{fig:phase_dia},
in the deep sub-Ohmic regime with $s$ below the critical point $s_{c}$, only the coherent phase exists. Above $s_{c}$, two lines of the
dynamical crossover separate the domain into three parts, and the coherent phase, the incoherent phase and the reemerged coherent phase, occur
sequentially with increasing coupling.

\begin{figure}[tbp]
  \centering
  \scalebox{0.35}{\includegraphics[clip]{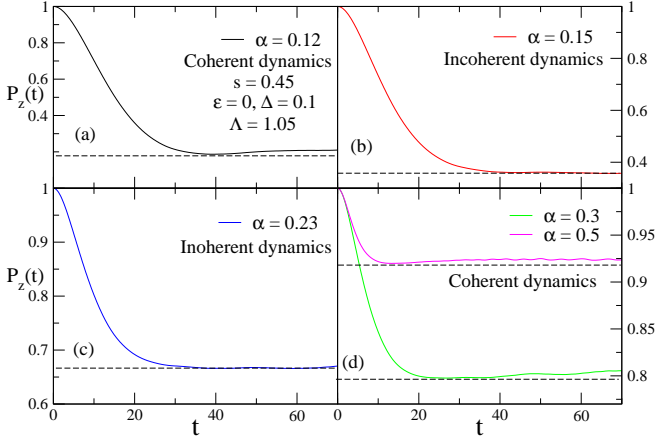}}
    \caption{The population difference $P_{z}(t)$ obtained by the multi-$\mathrm{D}_{1}$ \textit{Ansatz} for $s=0.45$ under the
      factorized bath initial condition for various coupling strengths $\alpha$.  Other parameters used are $\epsilon=0$, $\Delta=0.1$,
      $\Lambda=1.05$, and $N_{b}=230$.  (a) The coherent state occurs at $\alpha=0.13$. (b)-(c) The incoherent state occurs from
      $\alpha=0.15$ to $\alpha=0.23$. (d) The coherent state recurs at $\alpha=0.3$.  Dashed lines are plotted to show whether the dynamics
      is coherent or incoherent. }
    \label{fig:s0p45}
\end{figure}

\begin{figure}
  \scalebox{0.35}{\includegraphics[clip]{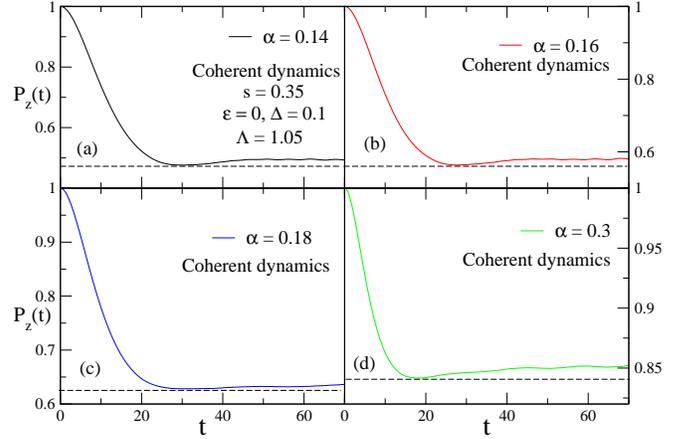}}
    \caption{The population difference $P_{z}(t)$ obtained by the multi-$\mathrm{D}_{1}$ \textit{Ansatz} for $s=0.35$ under the
      factorized bath initial condition for various coupling strengths $\alpha$.  Other parameters used are $\epsilon=0$, $\Delta=0.1$,
      $\Lambda=1.05$, and $N_{b}=230$.  (a)-(d) The coherent state persists at $\alpha=0.14$, $0.16$, $0.18$ and $0.3$.  Dashed lines are
      plotted to show the dynamics is coherent.}
    \label{fig:s0p35}
\end{figure}

To estimate the critical point $s_{c}$, simulations are performed for various values of $s$, corresponding to which dynamic behavior is
carefully evaluated.  In Figs.~\ref{fig:s0p45} and \ref{fig:s0p35}, horizontal lines are plotted at the minimum of $P_{z}(t)$ under
various values of $\alpha$ to illustrate whether the dynamics is coherent or incoherent.
Shown in Figs.~\ref{fig:s0p45}(a)-(d) is the population difference $P_{z}(t)$ at $s=0.45$ for $\alpha=0.12$, $0.15$, $0.23$, $0.3$ and
$0.5$.  A weak oscillation in $P_{z}(t)$ can be seen in Fig.~\ref{fig:s0p45}(a) for $\alpha=0.12$.
As demonstrated in Figs.~\ref{fig:s0p45}(b) and (c), the oscillations of $P_{z}(t)$ disappear and the system
moves to the incoherent phase within the range of $0.15\leqslant \alpha < 0.24$.  In this parameter region, the oscillation frequency tends
to $0$, and the damping rate increases with increasing $\alpha$.  The oscillation recurs for $\alpha> 0.24$, as plotted in
Fig.~\ref{fig:s0p45}(d) for $\alpha=0.3$ and $0.5$.  $P_{\rm{eq}}$, the steady-state value of $P_{z}(t)$, approaches $1$ and $P_{z}(t)$
reaches its minimum more rapidly, showing that the oscillation frequency and the damping rate increase with increasing coupling.  Contrary to
the case of $s=0.45$, sustained coherence is found for all coupling strengths if $s\leqslant 0.4$.   We perform dynamics simulations at
$s=0.35$ for $\alpha=0.14$, $0.16$, $0.18$ and $0.3$ to show there is no coherent-to-incoherent crossover for $s\leqslant 0.4$, which is
displayed in Figs.~\ref{fig:s0p35}(a)-(d).  With increasing coupling, both the damping rate and oscillation frequencies increase and only
the coherent phase can be seen. The result that the incoherent phase still exists at $s=0.45$, deviates from the NIBA prediction that
there is only the coherent phase for $s<0.5$.  The deviation may be attributed to that the NIBA method is not accurate for systems with long-lasting memory effects \cite{Kast13b}.

\begin{figure}[tbp]
  \centering
  \scalebox{0.35}{\includegraphics[clip]{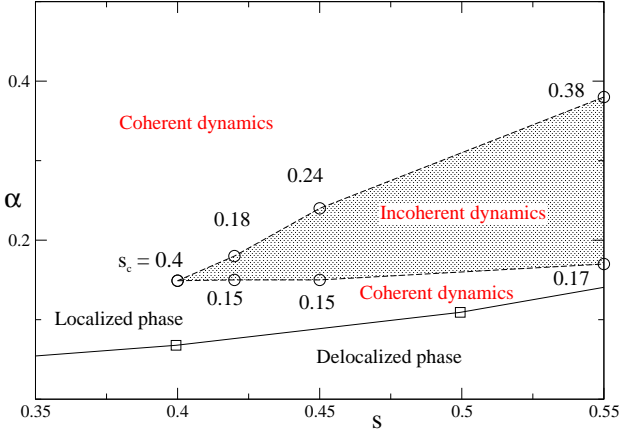}}
  \caption{The phase diagram of the dynamical coherent-incoherent crossover near the critical point $s_{c}$ obtained by the
    multi-$\mathrm{D}_{1}$ \textit{Ansatz}.  The critical point $s_{c}=0.4$ is estimated by the extrapolation.  The domain spanned by $s$ and
    $\alpha$ is separated by the dashed
    lines of the coherent-incoherent crossover.  The shaded area is the incoherent phase.  The
    solid line with square symbols is the localized-delocalized phase transition line and the data are extracted from Ref.~\cite{Winter}.}
  \label{fig:pha}
\end{figure}

Numerical results are carefully analyzed in the parameter space spanned by $s$ and $\alpha$ in order to arrive at a phase diagram in the
vicinity of $s_{c}$.  Simulations are performed for various coupling strengths $\alpha$ and four bath exponents $s=0.35$, $0.42$, $0.45$, and
$0.55$.  Whether the system is coherent or incoherent, is judged by the existence of population oscillations, which is illustrated
through Figs.~\ref{fig:s0p45} and \ref{fig:s0p35} with a step size of $\delta \alpha=0.01$.  A sketch of the phase diagram in the vicinity of
$s_{c}$, is shown in Fig.~\ref{fig:pha}.  The shaded incoherent phase domain is bordered by two lines, and with increasing $s$, the
vertical width of the domain expands.  Through extrapolation, we estimate $s_{c} \approx 0.4$.  Below $s_{c}$, there is
no incoherent state, as verified by our simulation for $s=0.35$.
To make the phase diagram in Fig.~\ref{fig:pha} comparable with the schematic one shown in Fig.~\ref{fig:phase_dia}, we extract data of
the localized-delocalized critical line from Ref.~\cite{Winter} and plot it with the black solid line with square symbols.  In the scope we plot, the
lower coherent-incoherent crossover line is above the phase transition line and their distance decreases with increasing $s$, which implies that the crossover line and the
phase transition line intersect at $s>0.5$.  The phase diagram, shown in Fig.~\ref{fig:pha}, is at variance with that obtained through
the NIBA method \cite{Kast13}.  From the NIBA method, the coherence does not recur in the strong coupling regime, thus the upper boundary of
the shaded area in Fig.~\ref{fig:phase_dia} does not exist in their phase diagram.  Meanwhile, the localized-delocalized phase transition
line is always
below the coherent-incoherent crossover line until they approximately meet at $s=0.5$.  The discrepancy of the two phase diagrams may be due
to the approximation used in the NIBA method.  In Ref.~\cite{Kast13}, two cases are analyzed.  For case of $s=0.5$, the authors employ the
condition $\alpha \ll \sqrt{\Delta/\omega_{c}} \approx 0.3$ for $\Delta=0.1$ and $\omega_{c}=1$.  While, the upper bound of the incoherent phase
for $s=0.5$, estimated as $\alpha\approx 0.3$ from Fig.~\ref{fig:pha}, is in the area of $\alpha \approx \sqrt{\Delta/\omega_{c}}$ in
which the NIBA method may not give an explicit prediction.
For $s \ll 1$, our
simulation regime is near $s=0.5$ where the NIBA method may be only accurate for short times \cite{Kast13_s}.
Though it is difficult to locate the crossover lines accurately from the numerical
simulations, it is our hope that the results still convincingly show that the coherence recurs for $s>s_{c}$ and
$s_{c} \approx 0.4$.   Recently, by analyzing the response function, the coherent-incoherent transition point $\alpha_{CI}$ and the critical exponent $s_{c}$ have been obtained in a slightly different setting \cite{Nalbach13}.  Furthermore, it is pointed out in Ref.~\cite{Nalbach13} that the critical $s_{c}$ decreases with decreasing $\Delta/\omega_{c}$.  The critical exponent obtained in this work should depend on $\Delta/\omega_{c}$ as well, and further simulations need to be carried out to clarify how $s_{c}$ varies with $\Delta/\omega_{c}$. 

\begin{figure}[tbp]
  \centering
  \scalebox{0.35}{\includegraphics[clip]{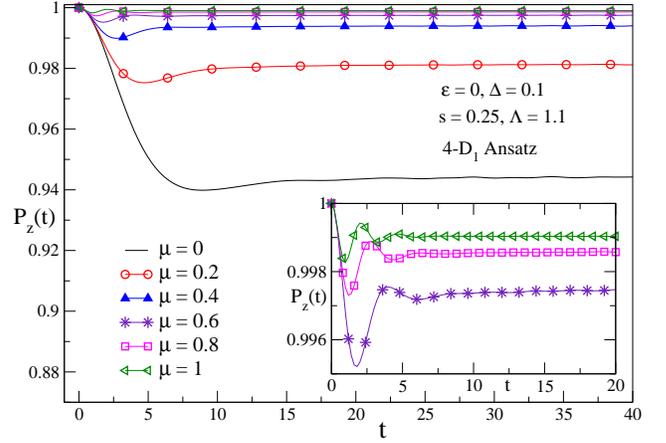}}
  \caption{The population difference $P_{z}(t)$ obtained by the multi-$\mathrm{D}_{1}$ \textit{Ansatz} under different initial bath
    preparation parameters $\mu$, $\mu=0$, $0.2$, $0.4$, $0.6$, $0.8$ and $1$. Other parameters used are $s=0.25$, $\Delta/\omega_{c}=0.1$,
    $\Lambda=1.1$ and $N_{b}=180$.}
  \label{fig:com_mu}
\end{figure}

To further confirm the picture of the dynamical crossover, $P_{z}(t)$ is investigated with different initial
baths.  The system is initialized in the ground state of
$\left. \sum_{l}\omega_{l}b_{l}^{\dagger}b_{l}+\frac{\sigma_{z}}{2}\sum_{l}\lambda_{l}\left(b_{l}^{\dagger}+b_{l}\right) \right|_{\sigma_{z}=\mu}$,
in which the parameter $\mu$ describes the polarization of the initial bath:  $\mu=0$ corresponds to
the factorized initial bath, and $\mu=1$ corresponds to the polarized initial bath.  For $\mu$ between $0$ and $1$, the bath is prepared
between the two limits.  As shown in Fig.~\ref{fig:com_mu}, $P_{z}(t)$ at $s=0.25$ under various values of $\mu$ ($\mu=0$, $0.2$,
$0.4$, $0.6$, $0.8$ and $1$) is plotted.  The steady-state value of $P_{z}(t)$ increases with increasing $\mu$.  As shown in the analysis
in Sec.~\ref{subsec:convergence}, the coupling term $\sigma_{z}/2\sum_{l}\lambda_{l}(b_{l}^{\dagger}+b_{l})$ in the Hamiltonian of
Eq.~(\ref{hamilton}), acts as a time-dependent bias $\epsilon(t)$.  As pointed out by Ref.~\cite{Nalbach10}, the system is in a
quasi-equilibrium state because of $\epsilon(t)$, and $P_{z}(t)$ appears as a steady-state value due to the
ultra-slow dynamics of the sub-Ohmic SBM.  Larger $\mu$ increases the initial value of $\epsilon(t)$ and yields larger $P_{z}(t)$.
Meanwhile, the frequency of the $P_{z}(t)$ is also increased because of the time-dependent bias.

\section{Conclusion}
\label{sec:CONCLUSION}

In this work, we have proposed an extended Davydov \textit{Ansatz}, called the multi-$\mathrm{D}_{1}$ \textit{Ansatz}, with a form analogous to
the Bloch wave function built from the Davydov $\mathrm{D}_{1}$ \textit{Ansatz} (previously known as the delocalized $\mathrm{D}_{1}$
\textit{Ansatz}).  Dynamics of the sub-Ohmic SBM has been investigated based on the multi-$\mathrm{D}_{1}$ \textit{Ansatz}.
We perform careful convergence tests for the logarithmic discretization parameter $\Lambda$ and the frequency lower bound $\omega_{\min}$.
To prevent artificial recurrence, a small $\Lambda$ should be employed.  With the same $\Lambda$, the dynamics with the factorized initial
bath is insensitive to $\omega_{\min}$, yet it is very sensitive with the polarized initial bath.
By quantifying how faithfully the {\it Ans\"atze} follows the Schr\"odinger equation through the relative
deviation, we show that the accuracy of the multi-$D_{1}$ \textit{Ansatz} is dramatically improved by increasing the multiplicity $M$.
The dynamical behaviors of the population difference $P_{z}(t)$ for different coupling strengths and initial bath conditions are
studied by the multi-$\mathrm{D}_{1}$ \textit{Ansatz}, with results consistent with those from the HEOM and PIMC approaches.  It is found that the
multi-$\mathrm{D}_{1}$ \textit{Ansatz} is a reliable yet unexpectedly efficient method that can treat the coherent and incoherent dynamics in a
unified manner.

The coherent-incoherent crossover is also investigated by the multi-$\mathrm{D}_{1}$ \textit{Ansatz}.  Based on the simulations for various
spectral exponents $s$, the critical point is estimated as $s_{c}\approx 0.4$, and the phase diagram near $s_{c}$ is obtained.
We find that for  $s<s_{c}$, the coherence will survive for all coupling strengths.  Above $s_{c}$, the system first goes from the coherent state to the
incoherent state with increasing $\alpha$, and when $\alpha$ increases further, the coherence recurs.

\section*{Acknowledgments}
The authors thank Vladimir Chernyak and Yiying Yan for useful discussion.
Support from the Singapore National Research Foundation through the Competitive Research Programme (CRP) under Project No.~NRF-CRP5-2009-04
is gratefully acknowledged.

\appendix
\section {The time dependent variational approach for the spin-boson model}
\label{sec:VAR_EOM}
In order to apply the Dirac-Frenkel time-dependent variational principle Eq.~(\ref{dirac_frenkel}), we first calculate the Lagrangian $L$ by
substituting Eq.~(\ref{Dsfun}) into Eq.~(\ref{Lag}),
\small{
\begin{eqnarray}
  L&=&\frac{i}{2}\sum_{n=1}^{M}\sum_{u=1}^{M} [A_{n}^{*}\dot{A}_{u} - \dot{A}_{n}^{*}A_{u} + \frac{1}{2} A_{n}^{*} A_{u} \sum_{k}
    (\dot{f}_{nk}f_{nk}^{*}  + f_{nk}\dot{f}_{nk}^{*} \nonumber\\
    && - \dot{f}_{uk}f_{uk}^{*} - f_{uk}\dot{f}_{uk}^{*}
    + 2 f_{nk}^{*}\dot{f}_{uk}
    - 2 \dot{f}_{nk}^{*}f_{uk} ) ] R\left( f_{n}^{*},f_{u}  \right) \nonumber \\
  &&+ \frac{i}{2}\sum_{n=1}^{M}\sum_{u=1}^{M} [B_{n}^{*}\dot{B}_{u} - \dot{B}_{n}^{*}B_{u}
    + \frac{1}{2} B_{n}^{*} B_{u}  \sum_{k} (\dot{g}_{nk}g_{nk}^{*} + g_{nk}\dot{g}_{nk}^{*} \nonumber\\
    &&- \dot{g}_{uk}g_{uk}^{*} - g_{uk}\dot{g}_{uk}^{*} + 2 g_{nk}^{*}\dot{g}_{uk}  - 2 \dot{g}_{nk}^{*}g_{uk} ) ]
  R\left( g_{n}^{*},g_{u}  \right) \nonumber\\
  &&-\Braket{\mathrm{D}(t)|\hat{H}|\mathrm{D}(t)},
  \label{Lagrangian}
\end{eqnarray}
}%
where $R(f_{n}, g_{m})$ is the Debye-Waller factor defined in Eq.~(\ref{debye_waller}), and the last term in Eq.~(\ref{Lagrangian}) with
$\epsilon=0$ can be obtained as
\small{
\begin{eqnarray}
    \label{aver_H}
  &&\Braket{\mathrm{D}(t)|\hat{H}|\mathrm{D}(t)} \nonumber \\
  &=&\sum_{n=1}^{M}\sum_{u=1}^{M}
  \left\{A_{n}^{*} A_{u} \left[ \sum_{k} \omega_{k} f_{nk}^{*}f_{uk}  \right.
    + \sum_{k}\frac{\lambda_{k}}{2}(f_{nk}^{*} + f_{uk}) \right] \nonumber\\
  && R\left( f_{n}^{*},f_{u}  \right) - \frac{\Delta}{2} A_{n}^{*}B_{u} R\left( f_{n}^{*},g_{u}  \right)\nonumber\\
  &&  +B_{n}^{*} B_{u} \left[  \sum_{k} \omega_{k} g_{nk}^{*}g_{uk}
    - \sum_{k}\frac{\lambda_{k}}{2}(g_{nk}^{*} + g_{uk}) \right] R\left( g_{n}^{*},g_{u} \right) \nonumber\\
  &&  - \frac{\Delta}{2} B_{n}^{*} A_{u} R\left( g_{n}^{*},f_{u}  \right) \Bigg\}.
\end{eqnarray}
}

Then, Dirac-Frenkel time-dependent variational principle yields the equations of motion for $A_{n}$ and $B_{n}$,
\small{
\begin{eqnarray}
  \label{eq_An}
  0&=&\sum_{m=1}^{M}\left\{
  \left[i\dot{A}_{m}- \frac{i}{2}A_{m}\sum_{k}
    \left(\dot{f}_{mk}f_{mk}^{*}+f_{mk}\dot{f}_{mk}^{*}-2f_{nk}^{*}\dot{f}_{mk}\right)\right] \right. \nonumber\\
  &&\left. +A_{m}\left[\sum_{k}\omega_{k}f_{nk}^{*}f_{mk} + \sum_{k} \frac{\lambda_{k}}{2}\left(f_{nk}^{*}+f_{mk}\right)\right] \right\}
  \nonumber\\
  &&R\left(f_{n}^{*},f_{m}\right) - \frac{\Delta}{2} \sum_{m=1}^{M}B_{m}R\left(f_{n}^{*},g_{m}\right),
\end{eqnarray}
}
and
\small{
\begin{eqnarray}
  \label{eq_Bn}
  0 &=& \sum_{m=1}^{M}\left\{ \left[i\dot{B}_{m} - \frac{i}{2}B_{m}
    \sum_{k} \left(\dot{g}_{mk}g_{mk}^{*}+g_{mk}\dot{g}_{mk}^{*}-2g_{nk}^{*}\dot{g}_{mk}\right)\right] \right. \nonumber\\
  &&\left. + B_{m}\left[\sum_{k}\omega_{k}g_{nk}^{*}g_{mk} - \sum_{k} \frac{\lambda_{k}}{2}\left(g_{nk}^{*}+g_{mk}\right)\right] \right\}
  \nonumber\\
  &&R\left(g_{n}^{*},g_{m}\right) - \frac{\Delta}{2} \sum_{m=1}^{M} A_{m} R\left(g_{n}^{*},f_{m}\right).
\end{eqnarray}
}
And the equations of motion for $f_{nl}$ and $g_{nl}$ are
\small{
\begin{eqnarray}
  \label{eq_fnl}
  0&=&\frac{i}{2}\sum_{m=1}^{M} \Bigg\{
  \left[2\dot{A}_{m}f_{ml}+2A_{m}\dot{f}_{ml}- A_{m}f_{ml} \right. \nonumber\\
    &&\left. \sum_{k}\left(\dot{f}_{mk}^{*}f_{mk}+f_{mk}^{*}\dot{f}_{mk}-2f_{nk}^{*}\dot{f}_{mk}\right)\right]
  +\omega_{l}A_{m}f_{ml} +\frac{\lambda_{l}}{2}A_{m} \nonumber\\
  &&\left. +A_{m}f_{ml}\left[\sum_{k}\omega_{k}f_{nk}^{*}f_{mk}
    + \sum_{k}\frac{\lambda_{k}}{2}\left(f_{nk}^{*}+f_{mk}\right)\right]\right\}
   \nonumber\\
  && R\left(f_{n}^{*},f_{m}\right) -\sum_{m=1}^{M}\frac{\Delta}{2}B_{m}g_{ml}R\left(f_{n}^{*},g_{m}\right),
\end{eqnarray}
}
and
\small{
\begin{eqnarray}
  \label{eq_gnl}
  0&=&\frac{i}{2}\sum_{m=1}^{M}\Bigg \{
  \left[2\dot{B}_{m}g_{ml}+2B_{m}\dot{g}_{ml}-B_{m}g_{ml} \right.\nonumber\\
    && \left. \sum_{k}\left(\dot{g}_{mk}^{*}g_{mk}+g_{mk}^{*}\dot{g}_{mk}-2g_{nk}^{*}\dot{g}_{mk}\right)\right]
  +\omega_{l}B_{m}g_{ml}-\frac{\lambda_{l}}{2}B_{m} \nonumber\\
  &&\left.+B_{m}g_{ml}\left[\sum_{k}\omega_{k}g_{nk}^{*}g_{mk}-\sum_{k}\frac{\lambda_{k}}{2}\left(g_{nk}^{*}+g_{mk}\right)\right]\right\}
   \nonumber\\
  && R\left(g_{n}^{*},g_{m}\right) - \sum_{m=1}^{M}\frac{\Delta}{2}B_{m}g_{ml}R\left(g_{n}^{*},f_{m}\right).
\end{eqnarray}
}

\section {Hierarchy equation of motion description of sub-Ohmic spin-boson model}
\label{sec:HEOM}
For the spin-boson model Eq.~(\ref{hamilton}), let us denote the eigenstate for the $\sigma_{z}$ as $\sigma$, then the reduced density matrix
element for the two-level system is expressed in the path integral form with the factorized initial condition as \cite{Feynman,Grabert}

\begin{eqnarray}
    \label{Rho}
    &&\rho(\sigma,\sigma^{'};t)= \int\mathcal{D}\sigma\int\mathcal{D}\sigma^{'}\rho(\sigma_{0},\sigma_{0}^{'};t_{0}) \nonumber \\
    &&\times{e^{iS[\sigma;t]}}F(\sigma,\sigma^{'};t)e^{-iS[\sigma^{'};t]}.
  \end{eqnarray}

Here, $S[\sigma]$ is the action of the two-level system, and $F[\sigma,\sigma^{'}]$ is the Feynman-Vernon influence functional given by
\begin{eqnarray}
    \label{IF}
    F(\sigma,\sigma^{'};t)&=&\exp\left(-\int_{0}^{\infty} \mathrm{d} \omega{J(\omega)}
    \int_{t_{0}}^{t} \mathrm{d}\tau \int_{t_{0}}^{\tau} \mathrm{d}\tau^{'}  \right.\nonumber\\
    &&V^{\times}(\tau)\times \bigg[V^{\times}(\tau^{'})\coth\left(\frac{\beta\omega}{2}\right)\cos(\omega(\tau-\tau^{'})) \nonumber\\
      &&-iV^{\circ}(\tau^{'})\sin(\omega(\tau-\tau^{'})) \bigg] \bigg).
  \end{eqnarray}
Here we have introduced the abbreviations
\begin{eqnarray}
V&=&\frac{\sigma_z}{2},\\
V^{\times}&=&V[\tau]-V[\tau^{'}],\\
V^{\circ}&=&V[\tau]+V[\tau^{'}].
\end{eqnarray}
The correlation function can be written as
\begin{equation}
  \label{CF}
  C(t)=\int_{0}^{\infty} \mathrm{d}\omega{J(\omega)}\left[\coth \left(\frac{\beta\omega}{2} \right)\cos(\omega{t})-i\sin(\omega{t})\right].
\end{equation}
Here, $\beta$ is the inverse of temperature.  If we consider sub-Ohmic spectral density in Eq.~(\ref{spectral_soft}) and zero temperature case, the correlation function can be easily obtained
\begin{eqnarray}
    \label{ZCF}
    C(t)&=&\int_{0}^{\infty} \mathrm{d}\omega {J(\omega)} [\cos{\omega{t}}-i\sin{\omega{t}}] \nonumber \\
    &=&2\alpha\omega_{c}^{1-s} \Bigg\{\frac{\Gamma(s+1)}{(t^{2}+\frac{1}{\omega_{c}^2})^{\frac{s+1}{2}}}[\cos((s+1)\arctan(\omega_c{t})) \nonumber\\
      &&-i\sin ((s+1)\arctan(\omega_{c}{t}) )] \Bigg\},
\end{eqnarray}
with $\Gamma(x)$ denotes Gamma function.  We can further fit the correlation function Eq.~(\ref{ZCF}) by a set of exponential function as
\begin{equation}
  \label{FitCF}
  C(t)=\sum_{k=1}^{\mathrm{NA}}a_ke^{-\gamma_kt}-i\sum_{k=1}^{\mathrm{NB}}b_ke^{-\nu_kt}.
\end{equation}
Thus, the influence functional Eq.~(\ref{IF}) can be expressed as
\begin{eqnarray}
  &&F(\sigma,\sigma^{'};t) = \nonumber\\
  &&\prod_{k=1}^{\mathrm{NA}}
  \exp \left(-\int_{t_{0}}^t \mathrm{d}\tau \int_{t_0}^{\tau} \mathrm{d}\tau^{'}V^{\times}(\tau)V^{\times}(\tau^{'})a_{k}e^{-\gamma_{k}(\tau-\tau^{'})} \right) \nonumber\\
  &&\times\prod_{k=1}^{\mathrm{NB}}
  \exp \left(-\int_{t_{0}}^{t} \mathrm{d}\tau\int_{t_{0}}^{\tau} \mathrm{d}\tau^{'}V^{\times}(\tau)V^{\circ}(\tau^{'})-ib_{k}e^{-\nu_{k}(\tau-\tau^{'})} \right). \nonumber\\
\end{eqnarray}
Taking the derivative of Eq.~(\ref{Rho}), we have
\begin{eqnarray}
  &&\frac{\partial}{\partial{t}}\rho(\sigma,\sigma^{'};t)
  =-i\mathcal{L}\rho(\sigma,\sigma^{'};t)-V^{\times}(t)
  \int\mathcal{D}\sigma\int\mathcal{D}\sigma^{'}\rho(\sigma_0,\sigma_0^{'};t_0)\nonumber\\
  &&\left[\int_{t_0}^{t} \mathrm{d}{\tau}V^{\times}(\tau)\sum_{k=1}^{\mathrm{NA}}a_ke^{-\gamma_k(t-\tau)}
      -i\int_{t_0}^{t} \mathrm{d}{\tau}V^{\circ}(\tau)\sum_{k=1}^{\mathrm{NB}}b_ke^{-\nu_k(t-\tau)} \right]  \nonumber\\
  && \times{e^{iS[\sigma,t]}}F(\sigma,\sigma^{'};t)e^{-iS[\sigma^{'};t]}.
\end{eqnarray}
In order to derive the equation of motion, we introduce the auxiliary operator
$\rho_{j_{1},\dots,j_{\mathrm{NA}};m_{1} ,\dots,m_{\mathrm{NB}}}(t)$
by its matrix element as \cite{Ishizaki,Tanimura}
 \begin{eqnarray}
   &&\rho_{j_1,...,j_{\mathrm{NA}};m_1,...,m_{\mathrm{NB}}}(\sigma,\sigma^{'};t)=\nonumber\\
   &&\int\mathcal{D}\sigma\int\mathcal{D}\sigma^{'}\rho(\sigma_0,\sigma_0^{'};t_0)
   \prod_{k=1}^{\mathrm{NA}} \left(\int_{t_0}^{t} \mathrm{d}{\tau}V^{\times}(\tau)a_{k}e^{-\gamma_k(t-\tau)} \right)^{j_k} \nonumber\\
    &&\prod_{k=1}^{\mathrm{NB}} \left(-i\int_{t_{0}}^{t} \mathrm{d}{\tau}V^{\circ}(\tau)b_{k}e^{-\nu_{k}(t-\tau)} \right)^{m_{k}} \nonumber \\
   &&\times{e^{iS[\sigma;t]}}F(\sigma,\sigma^{'};t)e^{-iS[\sigma^{'};t]},
 \end{eqnarray}
 for nonnegative integers $j_1$, $\dots$, $j_{\mathrm{NA}}$; $m_1$, $\dots$, $m_{\mathrm{NB}}$.  Note that only $\hat{\rho}_{0......0}(t)=\hat{\rho}(t)$ has a
 physical meaning and the others are introduced for computational purposes only.  Differentiating
 $\rho_{j_1,...,j_{\mathrm{NA}};m_1,...,m_{\mathrm{NB}}}(\sigma,\sigma^{'};t)$ with respect to $t$, we obtain the following hierarchy of
 equations in operator form
\begin{eqnarray}
  \label{HEOM}
  &&\frac{\partial}{\partial{t}}\hat{\rho}_{j_1,...,j_{\mathrm{NA}};m_1,...,m_{\mathrm{NB}}}(t) \nonumber\\
  &&=-\left [i\mathcal{L}+\sum_{k=1}^{\mathrm{NA}}j_k\gamma_k+\sum_{k=1}^{\mathrm{NB}}m_k\nu_k \right]
  \hat{\rho}_{j_1,...,j_{\mathrm{NA}};m_1,...,m_{\mathrm{NB}}}(t)\nonumber\\
  &&-V^{\times}(t) \left[\sum_{k=1}^{\mathrm{NA}}\hat{\rho}_{j_1,...,j_k+1,...j_{\mathrm{NA}};m_1,...,m_{\mathrm{NB}}}(t) \right. \nonumber\\
    && \left. + \sum_{k=1}^{\mathrm{NB}}\hat{\rho}_{j_1,...,j_{\mathrm{NA}};m_1,...,m_k+1,...m_{\mathrm{NB}}}(t) \right]\nonumber\\
&&+V^{\times}(t)\sum_{k=1}^{\mathrm{NA}}j_ka_k\hat{\rho}_{j_1,...,j_k-1,...j_{\mathrm{NA}};m_1,...,m_{\mathrm{NB}}}(t)\nonumber \\
&&-iV^{\circ}(t)\sum_{k=1}^{\mathrm{NB}}m_kb_k\hat{\rho}_{j_1,...,j_{\mathrm{NA}};m_1,...,m_k-1,...m_{\mathrm{NB}}}(t).
\end{eqnarray}

The HEOM consists of an infinite number of equations, but they can be truncated at finite number of hierarchy elements.  The infinite
hierarchy of Eq.~(\ref{HEOM}) can be truncated by the terminator as
\begin{eqnarray}
  && \frac{\partial}{\partial{t}}\hat{\rho}_{j_1,...,j_{\mathrm{NA}};m_1,...,m_{\mathrm{NB}}}(t)= \nonumber\\
  &&-\left[i\mathcal{L}+\sum_{k=1}^{\mathrm{NA}}j_k\gamma_k+\sum_{k=1}^{\mathrm{NB}}m_k\nu_k \right]
  \hat{\rho}_{j_1,...,j_{\mathrm{NA}};m_1,...,m_{\mathrm{NB}}}(t). \nonumber\\
\end{eqnarray}
The total number of hierarchy elements can be evaluated as $L_{\mathrm{tot}}=(N_{\mathrm{trun}}+NA+NB)!/N_{\mathrm{trun}}!(NA+NB)!$, while the total number of termination elements is $L_{\mathrm{term}}=(N_{\mathrm{trun}}+NA+NB-1)!/(NA+NB-1)!N_{\mathrm{trun}}!$, where $N_{\mathrm{trun}}$ is the depth of hierarchy for $m_{q\pm}(q=1,\cdots,N)$.  In practice, we can set the termination elements to zero and thus the number of hierarchy elements for calculation can be reduced as $L_{\mathrm{calc}}=L_{\mathrm{tot}}-L_{\mathrm{term}}$.

\end{document}